\documentclass[fleqn,usenatbib]{mnras}
\usepackage{newtxtext,newtxmath}
\usepackage[T1]{fontenc}
\usepackage{ae,aecompl}
\usepackage{graphicx}    
\title[Dark matter in BNS]{Effects of dark matter on the in-spiral properties of the binary neutron stars}
\author[H. C. Das et al.]{
H. C. Das$^{1,2}$
\thanks{E-mail: harish.d@iopb.res.in},
Ankit Kumar$^{1,2}$,
S. K. Patra$^{1,2}$,
\\
$^{1}$ Institute of Physics, Sachivalaya Marg, Bhubaneswar 751005, India\\
$^{2}$ Homi Bhabha National Institute, Training School Complex, 
Anushakti Nagar, Mumbai 400094, India\\}
\begin{document}
\maketitle
\date{\today}
\begin{abstract}
Using the relativistic mean-field model, we calculate the properties of binary neutron star (BNS) in the in-spiral phase. Assuming the dark matter (DM) particles are accreted inside the neutron star (NS) due to its enormous gravitational field, the mass $M$, radius $R$, tidal deformability $\lambda$ and dimensionless tidal deformability $\Lambda$ are calculated at different DM fractions. The value of $M$, $R$, $\lambda$ and $\Lambda$ decreases with the increase of DM percentage inside the NS. The in-spiral phase properties of the BNS are explored within the post-Newtonian (PN) formalism, as it is suitable up to the last orbits in the in-spiral phase. We calculate the strain amplitude of the polarization waveform $h_+$ and $h_\times$, (2,2) mode waveform $h_{22}$, orbital phase $\Phi$, frequency of the gravitational wave $f$ and PN parameter $x$ with DM as an extra candidate inside the NS. The magnitude of $f$, $\Phi$ and $x$ are almost the same for all assumed forces; however, the in-spiral time duration in the last orbit is different. We find that the BNS with soft equation of state and a high fraction of DM sustains more time in their in-spiral phase. We suggest that one should take DM inside the NS when they modelling the in-spiral waveforms for the BNS systems. 
\end{abstract}
\section{Introduction}\label{intro}
The detection of the binary neutron star (BNS) coalescence event GW170817 by LIGO/Virgo collaborations gives a new insight to study the compact objects  \citep{Abbott_2017,Abbott_2018}. The gravitational wave (GW) instrument observed this event about 40 Mpc distance from the source with a total mass of 2.74 $M_\odot$ of the binary \citep{Abbott_2017}. The gamma rays are detected after 1.7 seconds of the binary merger \citep{Abbott_multi2_2017,Goldstein_grb_2017}. In this set-up, the optical telescope identified an electromagnetic counterpart after 12 hours of the GW event near the NGC 4993 Galaxy called AT2017gfo \citep{Cowperthwaite_2017, Nicholl_2017, Chornock_2017}. The X-rays and radio waves are also detected several days after the merger \citep{Troja_2017, Margutti_2017, Hallinan_2017}. Hence, the BNS collision event witnesses the beginning of a new era in multi-messenger astronomy \citep{Abbott_multi1_2017}. The GWs coming from the in-spiral and merger of the compact binary stars are the most promising sources for ground-based detectors such as LIGO \citep{LIGO}, Virgo \citep{VIRGO} and KAGRA \citep{KAGRA}. The GWs encode information about the tidal deformability of the BNS system. The limits on the dimensionless tidal deformability parameter $\Lambda$ for the canonical star is given by GW170817 as $\Lambda_{1.4}=190_{-120}^{+390}$, which translates into a stringent bound $\Lambda\leq 580$ at the 90 \% confidence level \citep{Abbott_2018}. A considerable amount of research have been dedicated to constrain the mass $M$, radius $R$ and $\Lambda$ of the neutron star (NS) using the GW170817 data \citep{Margalit_2017,Radice_2018,Zhang_2018,Most_2018,Nandi_2019,Annala_2018}. Hence, the tidal deformability is the most promising parameter to constraint the EoSs, $M$ and $R$ of the NS. The detailed study of the GWs emitting from the in-spiral phase needs an accurate theoretical framework. This is because the tidal effects influence the dynamics of the in-spiral orbits and modify the corresponding gravitational waveforms. Substantial number of analysis have already been done to study the properties of BNS with tidal interactions as dominant contribution to the late in-spiral phase \citep{Flangnan_2008,Hinderer_2010,Vines_2011,Bini_2012,Read_2013,Favata_2014,Wade_2014,Agathos_2015}. 

The post-Newtonian (PN) formalism is suitable up to the last orbit to study the BNS systems in the in-spiral phase \citep{Blanchet_2006}. The high accuracy PN template for the in-spiral stage plays a key role in the analysis of GW data. This framework re-expresses the general relativistic description of particles into the standard equations of motion, which express in terms of acceleration of the particles \citep{Blanchet_2006}. 
The BNS in its in-spiral phase tidally interacts with each other, which affects their stellar structure. Each star's tidal property depends on the macroscopic quantity, such as mass $M$, radius $R$ and second Love number $k_2$,  which are all model-dependent in the evaluation \citep{Hinderer_2008}. Here, we used the relativistic mean-field (RMF) model to calculate the NS's properties. 

On the other hand, dark matter (DM) accounts for approximately 85\% of the matter in the Universe and about 27\% of its total mass-energy density. Theoretically, several DM particles have been hypothesised due to its unknown character such as  weakly interacting massive particle (WIMPs) \citep{Kouvaris_2011,Quddus_2020}, feebly interacting massive particles (FIMPs) \citep{Hall_2010, Bernal_2017}, Axions \citep{Duffy_2009} etc. The WIMP is the most abundant DM particle, which is also considered as the thermal relics of the Universe. It might have decayed in the dense region of the Universe, which produce the standard model (SM) particles, leptons, gamma-rays and neutrinos at freeze-out temperature \citep{Ruppin_2014}. Direct and indirect detection experiments have already been established to know the mystery of DM \citep{Bernabei_2008,  CDMS_2010, Xenon1T_2016}. Also, different models have been incorporated to study the effects of DM inside compact objects such as NS and white dwarf \citep{Bertone_2008, Kouvaris_2008, Kouvaris_2010}. The properties and origin of DM particles are not yet fully understood. In some formulation, it is treated as bosonic matter \citep{Bertoni_2013}. However, in maximum cases, the DM are fermions. In the present calculations, we consider Neutralino as a DM candidate, super-symmetric particle and the best candidate of WIMP \citep{Martin_1998, Panotopoulos_2017, Das_2019, Das_2020}. These particles interact with nucleons by exchanging SM Higgs boson. A detailed formalism and notations used in this paper can be found in Refs. \citep{Das_2020, Das_2021}.

The accretion of self-annihilating DM particles inside the NS affects the cooling properties of supernovae remnant \citep{Kouvaris_2008, Bhat_2019}. On the other hand, the non-annihilating DM particles develop interaction with baryons which affects the structure of the NS \citep{De_Lavallaz_2010, Ciarcelluti_2011}. Here, we assume that the WIMP is a DM particle that is non-annihilating in character. The addition of DM inside the NS generally reduces the macroscopic properties such as $M$, $R$ and $\lambda$ \citep{Das_2020, Quddus_2020, Das_2021}. This is because the EoS becomes softer with the increase of DM percentage. In the present case, we study the BNS properties in the in-spiral phase for DM admixed NS. The emission of GWs and its properties such as the (2,2) mode waveform $h_{22}$, polarization waveforms $h_+$ \& $h_\times$, GW frequency $f$, orbital phase $\Phi$ and PN parameter $x$ are calculated for equal mass binary NS. We take three different equation of states (EoSs) to see the parametric dependence, spanning from soft to stiff. All the in-spiral properties defined above are calculated in the retarded time interval.

The manuscript is organized as follow: In Sec. \ref{method}, we outline the basic RMF formalism required for the calculation of NS properties with the admixture of DM inside the NS. The in-spiral phase of the BNS in quasi-circular orbits are detailed in sub-sec. \ref{BNS}. In Sec. \ref{R&D}, we present our numerical results. Finally, we concluded our findings in Sec. \ref{con}.
\section{THEORETICAL FORMALISM}
\subsection{Equation of state}
\label{method}
In the extended RMF (E-RMF) formalism, the $\sigma-$meson is mainly responsible for the strong attraction at the intermediate range of the nuclear force, but the self-interaction of this scalar meson produces a weak repulsion at long-range \citep{Biswal_2015}. Because of the self-interaction of the $\sigma-$meson, the incompressibility of nuclear matter at nuclear saturation $K_{\infty}$ comes down to the empirical range of $K_{\infty}=210\pm{30}$ MeV \citep{BLAIZOT_1980}. The $\omega-$meson responsible for the strong hard-core repulsion of the nuclear potential and its self-interaction generates attraction at very short range and makes the nuclear EoS softer \citep{BODMER_1991, GMUCA_1992, Toki_1994, Bunta_2003}.  We take one RMF set NL3 \citep{Lalazissis_1997} and two E-RMF sets such as G3 \citep{Kumar_2017} and IOPB-I \citep{Kumar_2018} to calculate the NS properties in the present work.

Inside the NS, the neutron decays to proton, electron and anti-neutrino. For the stability of the system, the inverse $\beta$-decay takes place to maintain both $\beta$ equilibrium and charge neutrality. Hence, the EoS of NS is calculated from the contributions of both nucleons and leptons. The energy density and pressure of a NS is written as \citep{Das_2020, Das_2021}:
\begin{eqnarray}
{\cal {E}}_{NS} = {\cal {E}}_{NM}+ {\cal {E}}_{l}, 
\quad \mathrm{and} \quad P_{NS}= P_{NM}+ P_l.
\end{eqnarray}
Where ${\cal{E}}_{NM}$ and  $P_{NM}$ are the energy density and  pressure of the nucleon;  ${\cal{E}}_{l}$ and $P_{l}$ are for leptons, respectively.

After adding the DM to the NS, the total Lagrangian density is written as \citep{Das_2019,Das_2020}:
\begin{eqnarray}
{\cal{L}}& = & {\cal{L}}_{NS} + \bar \chi \left[ i \gamma^\mu \partial_\mu - M_\chi + y h \right] \chi +  \frac{1}{2}\partial_\mu h \partial^\mu h 
\nonumber \\
&&
- \frac{1}{2} M_h^2 h^2 + f \frac{M_{nucl.}}{v} \bar \psi h \psi , 
\label{eqdm}
\end{eqnarray}
where ${\cal{L}}_{NS}$ is the NS Lagrangian. The $\psi$ is the nucleon field and $\chi$ is the DM wave function. The parameters $y$, $f$ and $v$ are the
DM-Higgs coupling, nucleon-Higgs form factor and the vacuum expectation value of Higgs field, respectively. The values of $y$, $f$ and $v$ are 0.07, 0.35 and 246 GeV, respectively \citep{Das_2020, Das_2021}.

The two parameters $y$ and $f$ play significant role on the effect of DM in NS. Therefore, we constraint these two parameters using available experimental data. The expression  of scattering cross-section for the DM is written as \citep{Bhat_2019}:
\begin{equation}
\sigma_{SI}=\frac{y^2f^2M_{nucl.}^2}{4\pi}\frac{\mu_r}{v^2M_h^2},
\end{equation}
where $\mu_r$ is the reduced mass $\frac{M_{nucl.} \times M_\chi}{M_{nucl.}+M_\chi}$ and $M_\chi$ is the mass of the DM particle. The DM detection experiments, like XENON \citep{Xenon1T_2016}, PANDA-II \citep{PandaX_2016} and LUX \citep{LUX_2017} have put up their upper limit for the spin-independent scattering cross-section $\sigma_{SI}$ as $1.6\times10^{-47}$ cm$^{-2}$ ($M_{\chi}=50$ GeV), $1.6\times10^{-47}$ cm$^{-2}$ ($M_{\chi}=40$ GeV) and $1.6\times10^{-47}$ cm$^{-2}$ ($M_{\chi}=50$ GeV), respectively. The PANDA-II experiment has also put a limit on the mass of the WIMP as $M_{\chi}=5-1000$ GeV. The LHC has put a limit on the WIMP-nucleon scattering for Higgs portal DM in the range  $10^{-40}-10^{-50}$ cm$^{-2}$ \citep{PandaX_2016,Aad_2015}. We calculate the $\sigma_{SI}$ for three different masses of DM $M_{\chi}=50$, 100, and 200 GeV and their corresponding cross-sections are found to be 9.43, 9.60 and 9.70 in the order of ($10^{-46}$ cm$^2$) respectively. These data confirmed that the predicted values are consistent with the data within 90\% confidence level. This model also satisfies the LHC limit. The value of $f$ is reported in Ref. \citep{Djouadi_2012} using both lattice QCD \citep{Czarnecki_2010} and MILC results \citep{MILC_2009} whose value is $0.33_{-0.07}^{+0.30}$ \citep{Aad_2015} and our used $f$ (= 0.35) is well within the region.

The ${\cal{E}}$ and $P$ for NS with DM can be calculated by solving the Eq. (\ref{eqdm}) \citep{Das_2020, Das_2021}
\begin{eqnarray}
{\cal{E}}& = &  {\cal{E}}_{NS} + \frac{2}{(2\pi)^{3}}\int_0^{k_f^{DM}} d^{3}k \sqrt{k^2 + (M_\chi^\star)^2 } 
+ \frac{1}{2}M_h^2 h_0^2 ,
\label{etot}
\end{eqnarray}
and
\begin{eqnarray}
P& = &  P_{NS} + \frac{2}{3(2\pi)^{3}}\int_0^{k_f^{DM}} \frac{d^{3}k \hspace{1mm}k^2} {\sqrt{k^2 + (M_\chi^\star)^2}} 
- \frac{1}{2}M_h^2 h_0^2 ,
\label{ptot}
\end{eqnarray} 
where $k_f^{DM}$ and $M_\chi^*$ are the Fermi momentum and the effective mass of the DM respectively. $M_h$ is the mass of the Higgs equal to 125 GeV, and $h_0$ is the Higgs field calculated by applying the mean-field approximation \citep{Das_2019}. In the present calculations, the mass of the DM is taken as 200 GeV.

A large number of forces are available in the literature \citep{Dutra_2012,Kumar_2017,Kumar_2018}. Among them, NL3, G3 and IOPB-I are very much used both for finite nuclei as well as for infinite nuclear matter \citep{Kumar_2017,Kumar_2018}. The NL3 is one of the stiffest parameter set, which predicts a NS of  mass 2.775 $M_{\odot}$ and the G3 is the softest one with a predictive mass of 1.979 $M_{\odot}$. The values of the models are available in Ref. \citep{Kumar_2018}. Once the EoS has solved, the M-R profile of a single NS is calculated by solving the Tolmann-Oppenheimer-Volkoff (TOV) equation. When we consider a BNS system, the individual NS gets distorted due to the external field created by its companion. The tidal deformability parameter \big($\lambda=\frac{2}{3}k_2R^5$\big) created by the deformation depends on the structural/macroscopic properties. The dimensionless Love number $k_2$ can be calculated by solving two coupled differential equations as it is done in Refs. \citep{Hinderer_2008, Hinderer_2010, Kumar_2018}. The values of $R$ and $k_2$ are fixed for a given mass by the EoS for the NS. The dimensionless tidal deformability $\Lambda$ is defined as ($\Lambda=\lambda/M^5$).
\subsection{Binary neutron stars in quasi-circular orbits}
\label{BNS}
\subsubsection{Post-Newtonian expanded Taylor-T4}
\label{PN}
The PN formalism is a slow-motion, weakly stressed, and weak field approximation to the general theory of relativity and valid for any mass ratio \citep{Blanchet_2006, Buonanno_2014}. In the PN method, the equations of motion are obtained by systematically expanding the metric. The Einstein's field equations are expressed in powers of the dimensionless parameter $\epsilon=\sqrt{GM_t/c^2d}\sim\frac{v}{c}$, where $M_t$ is the total mass of the system, $d$ is the distance between two NS and $v$ is the characteristic velocity of particles. The metric is then solved iteratively in powers of $\epsilon$, and the equations of motion are evaluated from the metric using the geodesic equation. An expansion containing terms up to $\epsilon^n$ or equivalently $(1/c)^n$ is denoted an $\frac{n}{2}$PN expansion \citep{Blanchet_2006, Boyle_2007}. This calculation takes the energy and luminosity up to 3PN and 3.5PN order, respectively, for the in-spiralling binary in quasi-circular orbits. 

The BNSs emit GWs in their in-spiral phases. Because of the emission of the GWs for a classical system, the BNS have to rotate in quasi-circular orbits. The flux (luminosity) emitted by the system balances the rate of change of energy with time in that orbit, and the energy balance equation is given as \citep{Boyle_2007,Maggiore_2008,Lackey_2012}:
\begin{equation}
{\cal{L}}=-\frac{dE}{dt}=-\frac{dE/dx}{dt/dx},
\end{equation}
where $E$ is the energy of the system, $t$ is the time taken and $x$ is the PN parameter which is defined as \citep{Boyle_2007,Maggiore_2008,Lackey_2012}
\begin{equation}
x=\Big(M_t\frac{d\Phi}{dt}\Big)^{2/3}=\Big(M_t\Omega\Big)^{2/3},
\end{equation}
where $M_t$ is the total mass of BNS systems, $\Phi$ is the orbital phase, $\Omega=2\pi\omega$ is the orbital angular velocity ($\omega=f/2$  is the orbital angular frequency), and $f$ is the frequency of the emitted GW. The energy of the system calculated up to 3PN order in terms of $x$ is given as \citep{Blanchet_2006, Maggiore_2008, Lackey_2012}:
\begin{eqnarray}
E&=&-\frac{1}{2}M_t\nu x\Bigg\{1+\Bigg(-\frac{3}{4}-\frac{\nu}{12}\Bigg)x+\Bigg(-\frac{27}{8}+\frac{19\nu}{8}-\frac{\nu^2}{24}\Bigg)x^2
\nonumber \\
&&
+\Bigg[-\frac{675}{64}+\Bigg(\frac{34445}{576}-\frac{205\pi^2}{96}\Bigg)\nu
-\frac{155\nu^2}{96}-\frac{35\nu^3}{5184}\Bigg]x^3\Bigg\},
\label{enerPN}
\end{eqnarray}
where $\nu=m_1m_2/M_t^2$ is the symmetric mass ratio, $m_1$ and $m_2$ are the individual masses of the binary. The luminosity ${\cal{L}}$ is calculated by the time derivative of Eq. (\ref{enerPN}) using PN expansions up to 3.5 order is given as \citep{Blanchet_2006,Maggiore_2008,Lackey_2012}
\begin{eqnarray}
{\cal{L}}&=&\frac{32}{5}\nu^2x^5\Bigg\{1+\Bigg(-\frac{1247}{336}-\frac{35\nu}{12}\Bigg)x+4\pi x^{3/2}+\Bigg(-\frac{44711}{9072}
\nonumber\\
&&
+\frac{9271\nu}{504}+\frac{65\nu^2}{18}\Bigg)x^2
+\Bigg(-\frac{8191}{672}-\frac{583\nu}{24}\Bigg)\pi x^{5/2}
\nonumber\\
&&
+\Bigg[\frac{6643739519}{69854400}+\frac{16\pi^2}{3}-\frac{1712\gamma_E}{015}-\frac{856}{105}\ln(16x)
\nonumber\\
&&
+\Bigg(-\frac{134543}{7776}+\frac{41\pi^2}{48}\Bigg)\nu-\frac{94403\nu^2}{3024}-\frac{775\nu^3}{324}\Bigg]x^3
\nonumber\\
&&
+\Bigg(-\frac{16285}{504}+\frac{214745\nu}{1728}
+\frac{193385\nu^2}{3024}\Bigg)\pi x^{7/2}\Bigg\},
\label{lumPN}
\end{eqnarray}
where $\gamma_E\approx 0.5772$ is the Euler's constant. The phase evolution of the binary system can be calculated by solving the following equations:
\begin{eqnarray}
\frac{dx}{dt}=\frac{dE/dt}{dE/dx}=-\frac{{\cal{L}}}{dE/dx},
\label{dxdt}
\end{eqnarray}
\begin{eqnarray}
\frac{d\Phi}{dt}=\frac{x^{3/2}}{M_t}.
\label{dpdt}
\end{eqnarray}
There are various methods to integrate this system of equations labelled as TaylorT1-TaylorT4 \citep{Boyle_2007, Creighton_2011, Lackey_2012}.
In the TaylorT1 method, the Eqs. (\ref{enerPN}) and (\ref{lumPN}) are inserted in Eq. (\ref{dxdt}) and then the integration is performed by using the initial conditions $x_0=(M_t\Omega_0)^{2/3}$ and $\Phi_0$. In the TaylorT2 method, the equations are written by starting with a parametric solution of energy balance equations. Then, each expressions integrand is re-expanded as a single PN parameter x and truncated at the appropriate order \citep{Boyle_2007, Lackey_2012}:
\begin{eqnarray}
t(x)=t_0+\int_{x}^{x_0}dx\frac{(dE/dx)}{{\cal{L}}},
\label{tpn}
\end{eqnarray}
\begin{eqnarray}
\Phi(x)=\Phi_0+\int_{x}^{x_0}dx\frac{x^{3/2}}{M_t}\frac{(dE/dx)}{{\cal{L}}}.
\end{eqnarray}
In the present calculations, we take TaylorT4 approximation. In this method, the right-hand side of the Eq. (\ref{tpn}) is re-expanded as a single series and truncated at 3.5PN order before doing the integration. After calculating $x$, $t$ and $\Phi$, one can know the properties related to the in-spiralling binaries for the point-particle system. 

In the case of BNS, the tidal interaction comes in to picture, which has a significant role in the in-spiral properties. Therefore, we have to include the extra part of tidal interactions with this point-particle approximation. The motion of the system is tidally interacting, which influences their internal structures. Also, the tidal interactions affect the evolution of the GW phase by a parameter $\lambda$. Hence, the BNS system's tidal contributions must be added on the right-hand side of Eq. (\ref{dxdt}), which is modified as \citep{Baiooti_2011,Hotokezaka_2013,Hotokezaka_2016}:
\begin{equation}
\frac{dx}{dt}=\frac{64\nu}{5M_t} x^5\{F_{3.5}^{\mathrm{Taylor}}(x)+F^{\mathrm{Tidal}}(x)\},
\end{equation}
where $F_{3.5}^{\mathrm{Taylor}}$is PN-expanded expression for point-mass contribution using Taylor-T4 approximation, given by \citep{Blanchet_2008,Boyle_2007,Baiooti_2011,Hotokezaka_2013,Hotokezaka_2016}
\begin{eqnarray}
&&F_{3.5}^{\mathrm{Taylor}}(x)=1-\Bigg(\frac{743}{336}+\frac{11}{4}\nu\Bigg)x+4\pi x^{3/2}+\Bigg(\frac{34103}{18144}+\frac{13661}{2016}\nu
\nonumber\\ 
&&
+\frac{59}{18}\nu^2\Bigg)x^2-\Bigg(\frac{4159}{672}+\frac{189}{8}\nu\Bigg)\pi x^{5/2}
+\Bigg[\frac{16447322263}{139708800}-\frac{1712}{105}\gamma_E
\nonumber\\ 
&&
-\frac{56198689}{217728}\nu+\frac{541}{896}\nu^2-\frac{5605}{2592}\nu^3+\frac{\pi^2}{48}(256+451\nu)
\nonumber\\ 
&&
-\frac{856}{105}\ln(16x)\Bigg]x^3 +\Bigg(-\frac{4415}{4032}+\frac{358675}{6048}\nu+\frac{91495}{1512}\nu^2\Bigg)\pi x^{7/2},
\end{eqnarray}
With the addition of the tidal part to the PN formalism, we adopt the method developed by Vines {\it et al.} \citep{Vines_2011} up to 1PN accuracy, which is expressed as 
\begin{eqnarray}
&F^{\mathrm{Tidal}}(x)&=\frac{32\chi_1\lambda_2}{5M_t^6}\Big[12(1+11\chi_1)x^{10}+\Big(\frac{4421}{28}-\frac{12263}{28}\chi_2
\nonumber\\ 
&&
+\frac{1893}{2}\chi_2^2-661\chi_2^3\Big)x^{11}
+(1\leftrightarrow2)\Big],
\label{tidalEq}
\end{eqnarray}
where $\lambda_2$ is the tidal deformability of the second star and $\chi_1$ is the mass fraction of the first star ($\chi_1=M_1/M_t$). The numerical values of $k_2$ and $R$ are given in Table \ref{table1} for different parameter sets with three DM fractions. The numerical values are different from others because they are model dependent. 
Baiotti {et al.} \citep{Baiooti_2011} re-expressed the Eq. (\ref{tidalEq}) in terms of tidal Love number $k_2$ and compactness parameter $C$  as
\begin{equation}
F^{\mathrm{Tidal}}(x)=\sum_{I=1,2}F_{LO}(\chi_I)x^5\big(1+F_1(\chi_I)x\big),
\end{equation}
with 
\begin{equation}
F_{LO}(\chi_I)=4\hat{k_2^I}\Bigg(\frac{12-11\chi_I}{\chi_I}\Bigg),
\end{equation}
and 
\begin{equation}
F_1(\chi_I)=\frac{4421-12263\chi_I+26502\chi_I^2-18508\chi_I^3}{336(12-11\chi_I)}.
\end{equation}
The expression for $\hat{k_2^I}$ is defined as:
\begin{equation}
\hat{k_2^I}=k_2^I\Bigg(\frac{\chi_I}{C_I}\Bigg)^5  \ \ I=1,2.
\end{equation}
For equal-mass binary, $\chi_1 = \chi_2=\chi=1/2$ and $C=C_1=C_2$, the Eq. (\ref{tidalEq}) can be written as \citep{Hotokezaka_2013}  
\begin{equation}
  F^{\mathrm{Tidal}}(x)= \frac{52}{5M_t}\frac{k_2}{C^5}x^{10}\Big(1+\frac{5203}{4368}x\Big).
\end{equation}
Although the tidal interactions affect the NS-NS in-spiral only at 5PN order, its coefficient is 10$^4$ order magnitude for a NS with a radius of 10–15 km, $k_2\sim0.1$, and $C\sim0.14–0.20$. Hence, it plays an important role in the late in-spiral stage.
\subsubsection{Polarization waveforms}
\label{pola}
From the observational point of view, the GW detector detects the radiation in the direction of source. Therefore, the polarization waveforms are measured in the line-of-sight of source, which can be expressed in spherical co-ordinates ($R,\hat{\theta},\hat{\phi}$) as \citep{Maggiore_2008,Isoyama_2020}:
\begin{eqnarray}
&&h_+(\mathrm{t})=\frac{4}{r}{\cal{M}}^{5/3}\omega^{2/3}\Bigg(\frac{1+\cos^2\theta}{2}\Bigg)\cos(2\omega t_{ret}+2\phi),\nonumber\\ \mathrm{and} \\ \nonumber
&&
h_\times(\mathrm{t})=\frac{4}{r}{\cal{M}}^{5/3}\omega^{2/3}\cos\theta\sin(2\omega t_{ret}+2\phi),
\label{h+hx}
\end{eqnarray}
where the chirp mass ${\cal{M}}$ is defined as ${\cal{M}}=\nu^{3/5}M$. The amplitudes of the GW polarization with a fixed $\omega$ depends on the binary masses through ${\cal{M}}$, which can also be derived as ${\cal{M}}=\frac{(m_1m_2)^{3/5}}{(m_1+m_2)^{1/5}}$. It is measured by LIGO with a good precision ${\cal{M}}=1.188_{-0.002}^{+0.004}M_\odot$ \citep{Abbott_2017}. If we see the orbit edge-on, $\theta=\pi/2$, then $h_\times$ vanishes and the GW is linearly polarized. If we put, $\theta=0$, then $h_+$ and $h_\times$ have the same amplitude, but only difference is the phase and hence it is circularly polarized.

The polarization waveforms for an equal mass binary NS (EMBNS) systems along the z-axis (optimally oriented observer) calculated up to 2.5PN are given by  \citep{Arun_2004, Arun_2005, Kidder_2007, Boyle_2007}
\begin{eqnarray}
h_+^{(z)}&=&\frac{M_t}{2D}x\Bigg(\cos2\Phi\bigg\{-2+\frac{17}{4}x-4\pi x^{3/2}+\frac{15917}{2880}x^2+9\pi x^{5/2}\bigg\}
\nonumber\\ 
&&
+\sin2\Phi\Big\{\frac{59}{5}x^{5/2}\Big\}\Bigg),
\end{eqnarray}
\begin{eqnarray}
h_\times^{(z)}&=&\frac{M_t}{2D}x\Bigg(\sin2\Phi\bigg\{-2+\frac{17}{4}x-4\pi x^{3/2}+\frac{15917}{2880}x^2+9\pi x^{5/2}\bigg\}
\nonumber\\ 
&&
+\cos2\Phi\Big\{-\frac{59}{5}x^{5/2}\Big\}\Bigg).
\end{eqnarray}
The polarization waveforms are shown in the Fig. \ref{hphc} for IOPB-I parameter sets. 
The dominant (2,2) mode up to 3PN order is written as follow \citep{Blanchet_2008,Kidder_2008, Lackey_2012}
\begin{eqnarray}
&h_{22}&=-8\nu\sqrt{\frac{\pi}{5}}\frac{M_t}{D}e^{-2i\Phi}x  \Bigg \{1+\Bigg(-\frac{107}{42}+\frac{55\nu}{42}\Bigg)x+2\pi x^{3/2}
\nonumber\\
&&
+\Bigg(-\frac{2173}{1512}-\frac{1069\nu}{216}+\frac{2047\nu^2}{1512}\Bigg)x^2 
+\Bigg[\frac{-107\pi}{21}+\Bigg(\frac{34\pi}{21}
\nonumber\\
&&
-24i\Bigg)\nu\Bigg]x^{5/2}+\Bigg[\frac{27027409}{646800}-\frac{856\gamma_E}{105}+\frac{2\pi^2}{3}+\frac{428i\pi}{105}
\nonumber\\
&&
-\frac{428}{105}\ln(16x)
+\Bigg(\frac{41\pi^2}{96}-\frac{278185}{33264}\nu\Bigg)-\frac{20261\nu^2}{2772}
\nonumber\\
&&
+\frac{11463\nu^3}{99792}\Bigg]x^3 \Bigg \},
\end{eqnarray}
where $D$ is the distance between source and observer taken as 100 Mpc in our calculations. 
\section{RESULTS AND DISCUSSIONS}
\label{R&D}
In this section, we discussed our calculated results with three different parameter sets. On top of the baryonic and leptonic EoS, the contribution of DM on it and subsequently its effects on M-R profile, tidal deformability and GW properties are analyzed. The results are given in Table \ref{table1} and Figs. \ref{eos}-\ref{all2}.

\subsection{EOS, mass-radius relation, tidal deformability}
\label{NSprop}
\begin{figure}
\centering
\includegraphics[width=0.5\textwidth]{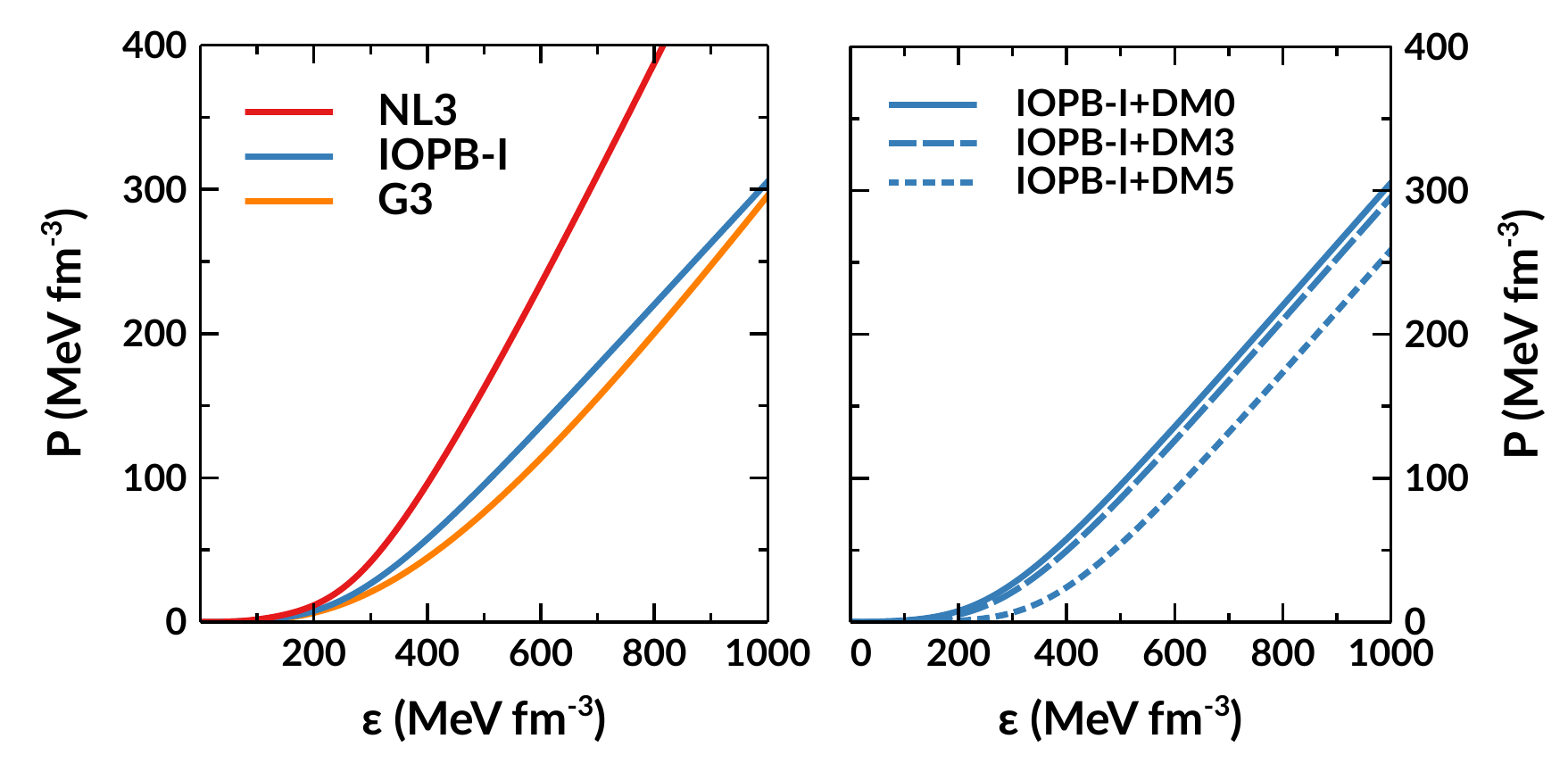}
\caption{(colour online) {\it Left}: EoSs are shown for 3-parameter sets NL3, IOPB-I and G3. {\it Right}: The EOSs are shown with the addition of DM with $k_f^{DM}$= 0, 0.03 and 0.05 GeV for IOPB-I parameter set as representative case.}
\label{eos}
\end{figure}

As it is stated earlier, the three RMF EoS, namely NL3 (stiff) \citep{Lalazissis_1997}, IOPB (less stiff) \citep{Kumar_2018} and the G3 (soft) \citep{Kumar_2017} for the core part are calculated. The well known BCPM EoS \citep{BKS_2015} is used uniformly for all these sets to include the effect of the lower dense regions, i.e. the crust part and the NS properties are calculated with these unified EOS. We add the DM inside the NS, which interacts with nucleons via SM Higgs (see Sub-Sec. \ref{method}). As a representative case, the EoS of  IOPB-I parameter set is chosen with DM momentum, $k_f^{DM}$= 0, 0.03 and 0.05 GeV, which in short, written as IOPB-I+DM0\footnote{same as IOPB-I because DM0 means no DM inside the NS}, IOPB-I+DM3, and IOPB-I+DM5 respectively. The NS properties such $R_{1.35}$, $\lambda_{1.35}$, $\Lambda_{1.35}$, $k_{2,1.35}$ are given in Table \ref{table1}. The predicted mass-radius for DM admixed NS with different DM momenta are also given for comparison. The maximum mass decreases with the increase of $k_f^{DM}$. The total EoS of the NS is depicted in Fig. \ref{eos}. From the right panel of Fig. \ref{eos}, it is clear that with the increase of DM momentum, the EoS becomes softer, which is consistent with our previous results \citep{Das_2020}.
\begin{figure}
\centering
\includegraphics[width=0.4\textwidth]{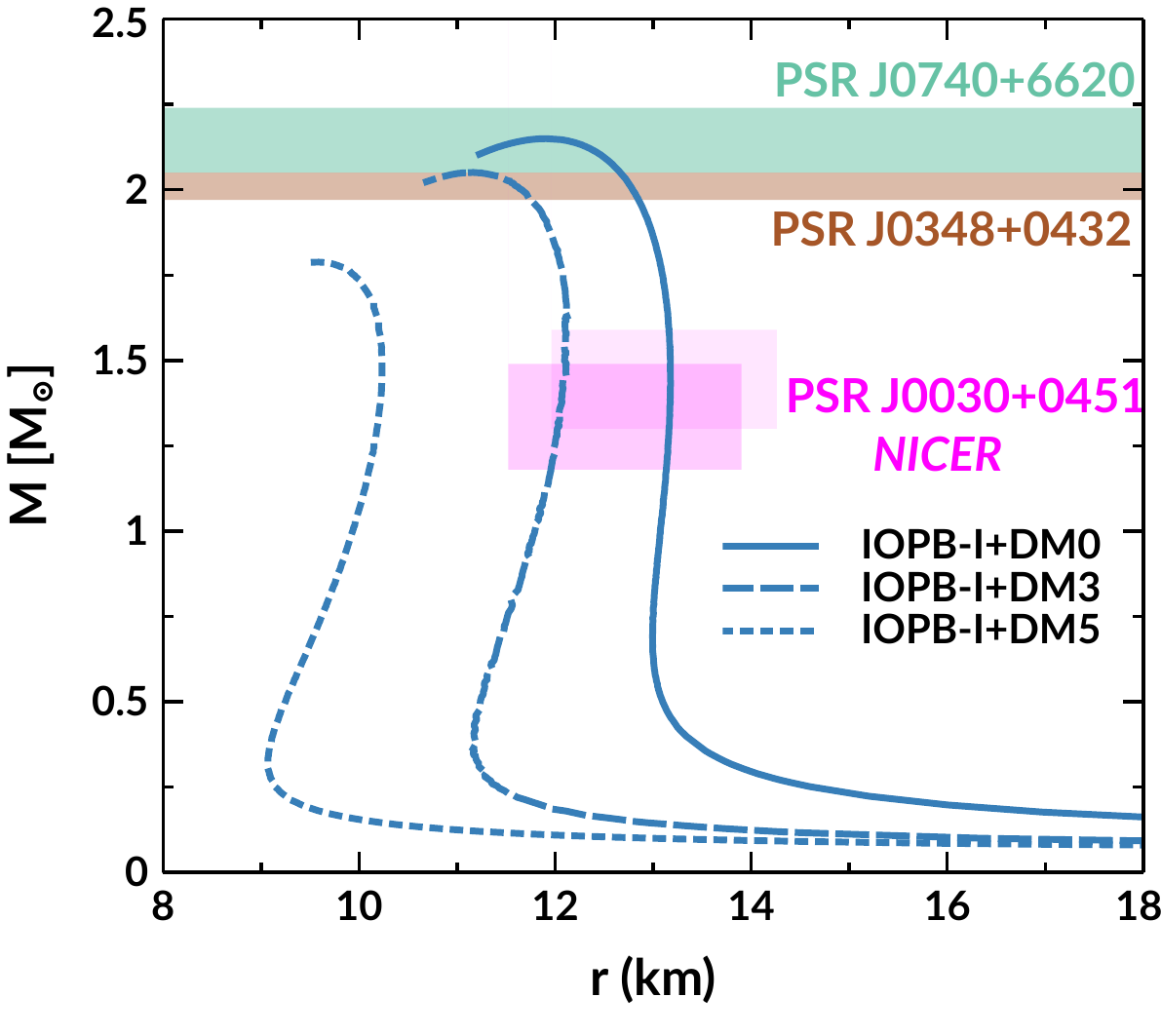}
\caption{(colour online) Mass-Radius plot for IOPB-I parameter sets with $k_f^{DM}$ is equal to 0, 0.03 and 0.05 GeV. Recently measured heavy NS, PSR J0740+6620 \citep{Cromartie_2019} and PSR J0348+0432 \citep{Antoniadis_2013} are shown with horizontal bars. The NICER results are also shown with rectangles from two different analysis \citep{Miller_2019, Riley_2019}.}
\label{mr}
\end{figure}

The $M$-$R$ relation of a star is calculated by solving the TOV equations with EoS as the main input with the boundary conditions $P=P_c$ at $r=0$ and $P=0$ at $r=R$. The masses and radii with IOPB-I parameter sets are calculated for DM admixed NS, which are shown in Fig. \ref{mr}. Recently measured massive pulsar PSR J0740+6620 \citep{Cromartie_2019}, PSR J0348+0432 \citep{Antoniadis_2013} and NICER \citep{Miller_2019,Riley_2019} data are also depicted. With the addition of DM, the maximum mass $M_{max}$ and radius $R_{max}$ decrease. For $k_f^{DM}$=0.03 GeV, the $M_{max}$ satisfy the constraint given by Cromartie {\it et al.} \citep{Cromartie_2019} as well as Antoniadis {\it et al.} \citep{Antoniadis_2013} and the radius constraint given by NICER \citep{Miller_2019,Riley_2019}. However, with $k_f^{DM}$= 0.05 GeV, the $M_{max}$ and $R_{max}$ become very less, which doesn't satisfy both massive pulsar and NICER data. Therefore, one can constraints the DM percentage inside the NS with the help of the observational data.

\begin{figure}
\centering
\includegraphics[width=0.5\textwidth]{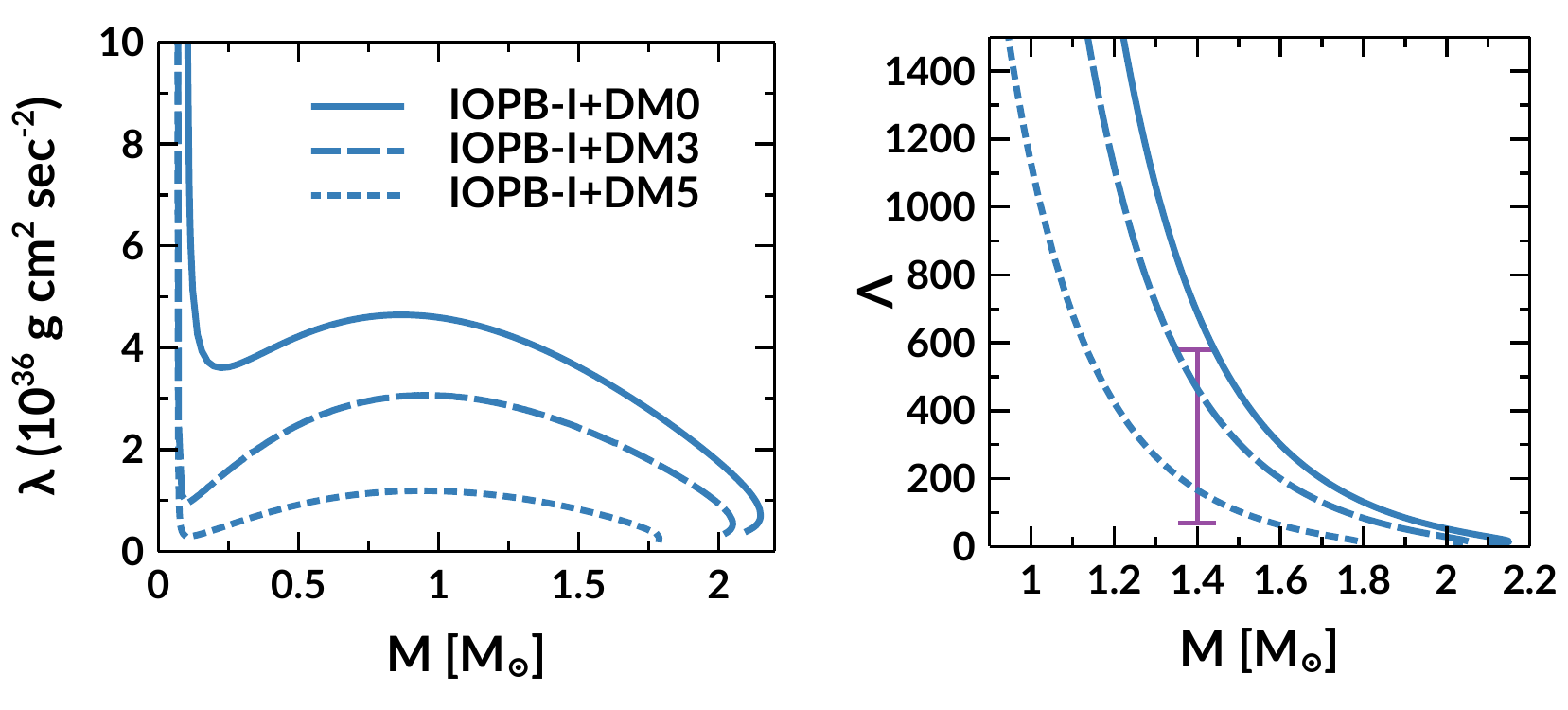}
\caption{(colour online) {\it Left}: The mass variation of tidal deformability of the NS is shown for IOPB-I set with two different DM momentum. {\it Right}: The dimensionless tidal deformability vs. masses of the NS with the addition of DM are shown. The vertical line with bar represents the GW170817 constraints on $\Lambda_{1.4}$ \citep{Abbott_2017,Abbott_2018}.}
\label{lambda}
\end{figure}

The tidal deformability of the NS decreases with DM's addition, as shown in Fig. \ref{lambda}. The same trend also obtained for $\Lambda$. This is because, with the addition of DM, the EoS becomes softer and the mass and radius change accordingly, as shown in Fig. \ref{mr}. Therefore, the softer EoS gives lower $\lambda$ as compare to the stiffer one. The vertical double-headed line represents the constraints from GW170817 with $\Lambda_{1.4}=190_{-120}^{+390}$ \citep{Abbott_2018}. The values of $\Lambda_{1.4}$ are 473.22, 169.644 for DM momentum 0.03 and 0.05 GeV, respectively. This implies that the value of $\Lambda_{1.4}$ with these DM momenta satisfies the constraints given by GW170817.
\begin{table*}
\centering
\caption{The maximum mass of the spherical NS $M_{max}$ obtained with different EOS, the radius $R_{1.35}$, the tidal deformability $\lambda_{1.35}$, the dimensionless tidal deformability $\Lambda_{1.35}$, the tidal love number $k_{2,1.35}$, the compactness $C_{1.35}$ for NS mass 1.35 $M_\odot$. The total mass of the BNS system's is $M_t=2.7 M_{\odot}$ and $M_t\Omega_0=M_t\pi f_0=0.0155$ are taken in this calculation with the minimum gravitational wave frequency $f_0$  $\approx$ 371 Hz. $\lambda_{1.35}^*$ is in $10^{36}$ g cm$^2$ sec$^{-2}$ unit.}
\label{table1}
\begin{tabular}{|llllllll|}
\hline \hline
EOSs & $M_{max}$ ($M_{\odot})$ & $R_{1.35}$ (km) & $\lambda_{1.35}^*$& $\Lambda_{1.35}$ & $k_{2,1.35}$ & $C_{1.35}$ & $M_t\Omega_0$ \\ \hline
NL3        & 2.775 & 14.568 & 7.415  & 1575.239 & 0.1130  & 0.137 & 0.0155 \\ \hline
G3         & 1.979 & 12.476 & 2.754  & 585.182  & 0.0911  & 0.159 & 0.0155 \\ \hline
IOPB-I     & 2.149 & 13.168 & 4.026  & 855.304  & 0.1017  & 0.151 & 0.0155 \\ \hline
IOPB-I+DM3 & 2.051 & 12.047 & 2.717  & 577.237  & 0.1071  & 0.165 & 0.0155 \\ \hline
IOPB-I+DM5 & 1.788 & 10.204 & 0.990  & 210.387  & 0.0896  & 0.195 & 0.0155 \\ \hline \hline
\end{tabular}
\end{table*}
\subsection{Gravitational waves properties}
\label{GW}
In this sub-sec., we describe the BNS's in-spiral properties using the PN formalism applied to quasi-circular orbit. For simplicity, we restrict our calculations only to EMBNS. Again, we take IOPB-I as a representative EoS for the DM admixed NS since it predicts the mass of the NS 2.149 $M_{\odot}$ which is consistent with PSR J0740+6620 \citep{Cromartie_2019}. Here, we calculate the polarization waveforms and the GW properties such as amplitude, frequency, orbital phase for the EMBNS. The observables are calculated at retarded time defined by \citep{Hotokezaka_2013}: 
\begin{eqnarray}
t_{ret}=t-r_*,
\label{ret1}
\end{eqnarray}
where $r_*$ is the tortoise coordinate defined as 
\begin{eqnarray}
r_*=r_A+2M_t\ln\Big(\frac{r_A}{2M_t}-1\Big),
\end{eqnarray}
where $r_A=\sqrt{A/4\pi}$ and A is the proper surface area of the sphere. In a simplify manner, $r_A$ is written as  \citep{Hotokezaka_2013}
\begin{eqnarray}
r_A=r\Bigg(1+\frac{M_t}{2r}\Bigg)^2,
\end{eqnarray}
where $r\approx200M_t$ is the finite spherical-coordinate radius.
\begin{figure}
\centering
\includegraphics[width=0.5\textwidth]{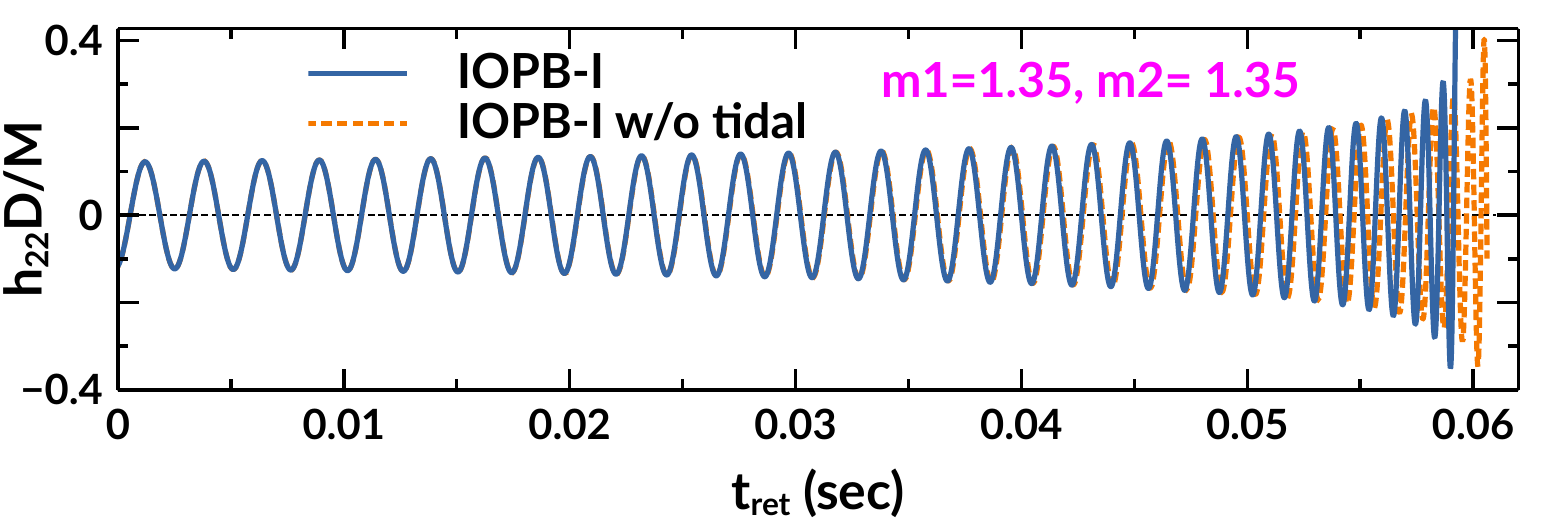}
\caption{(colour online) (2,2) mode waveforms are shown with and without tidal contributions for the IOPB-I parameter sets as a representative case.}
\label{iopb_wo_tidal}
\end{figure}
\begin{figure}
\centering
\includegraphics[width=0.5\textwidth]{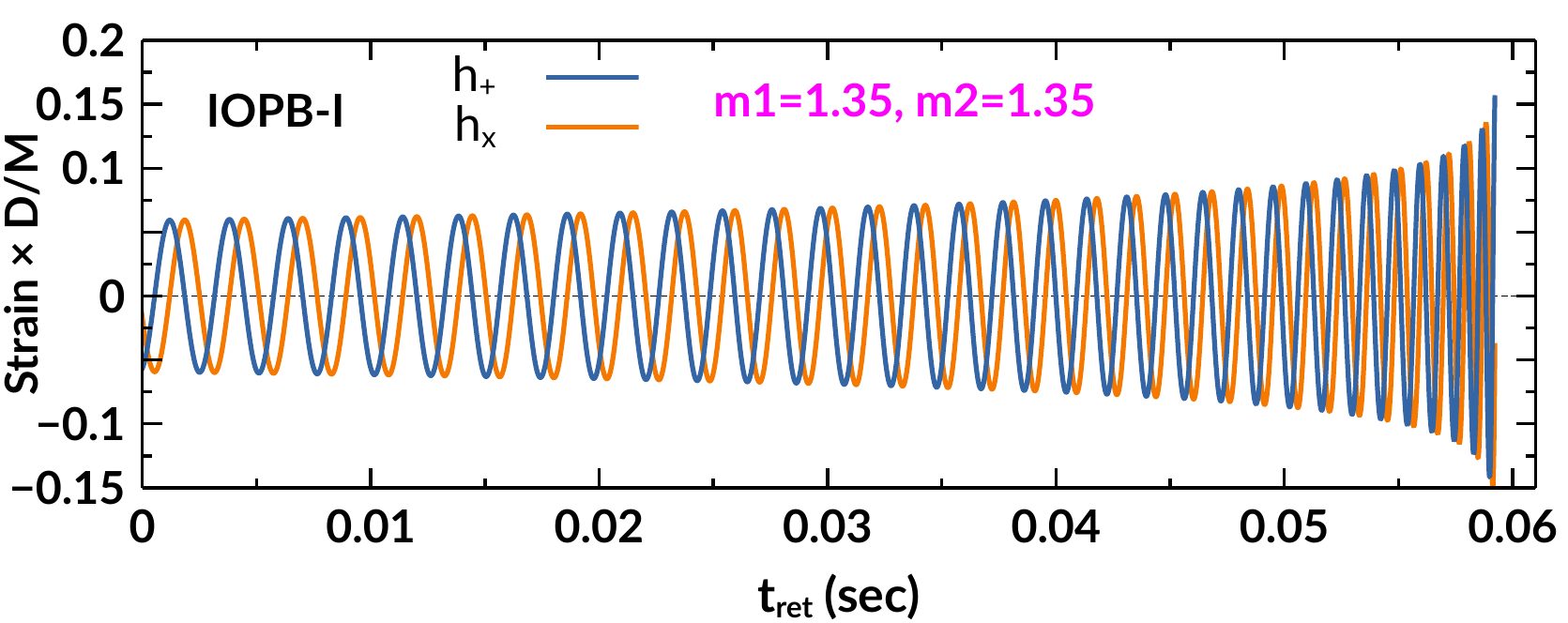}
\caption{(colour online) Plus and cross polarization waveforms are shown for the IOPB-I parameter sets.}
\label{hphc}
\end{figure}
\begin{figure}
\centering
\includegraphics[width=0.5\textwidth]{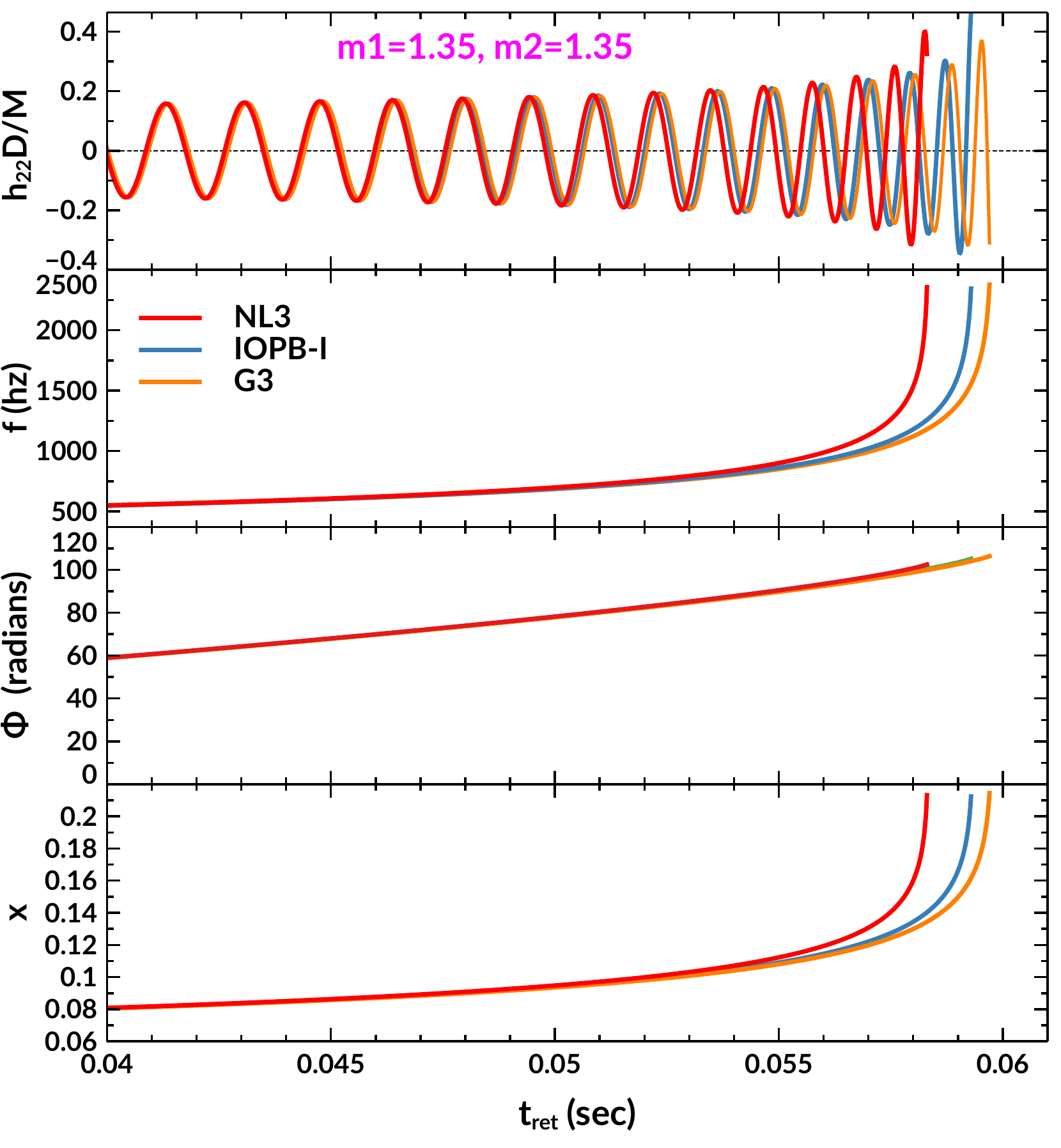}
\caption{(colour online) (2,2) mode waveform, frequency, phase and PN-parameters are shown for three parameter sets of the EMBNS in the retarded time interval calculated for $D=100$ Mpc.}
\label{all1}
\end{figure}

Fig. \ref{all1} shows the waveforms, frequencies, phases, and $x$ for NL3, G3 and IOPB-I sets. The stiff EoS, like NL3, predicts higher mass and tidal deformability as compared to the softest EoS like G3. Therefore, the massive NS easily deformed by the presence of tidal fields created by its companion star. Hence, a less massive BNS system sustains a longer time than a more massive BNS system in the in-spiral phase. This is because the tidal interactions accelerate the orbital evolution in the late in-spiral phases due to increased interactions between two NSs. Some significant effect has seen in Fig. \ref{iopb_wo_tidal} with and without tidal interactions.

To have a quantitative and comparative estimation for the in-spiral properties of the BNS merger, we enumerate the numerical value of $t_{ret}$, $\Phi$, $f$ and $x$ in Table \ref{table2} using NL3, IOPB-I and G3. The last orbital time period is 0.0583 sec. for NL3, which is less than from other two sets and confirms the model-dependent prediction. There is significant influence seen for $t_{ret}$, $\Phi$ and $f$, except the value of $x$. The frequency and phase are also increased for the BNS of soft EoS. The pictorial representations of $f$, $\Phi$ and $x$ are compared in Fig. \ref{all1} with and without DM contains.
\begin{table}
\centering
\caption{The orbital phase ($\Phi$), gravitational wave frequency ($f$), post-Newtonian parameter ($x$) are given in the last in-spiral orbits of the BNS system.}
\label{table2}
\begin{tabular}{|lllll|}
\hline \hline
EOSs & $t_{ret}$ (sec.) & $\Phi$ (radians)& $f$ (hZ) & \quad $x$ \\ \hline
NL3        & 0.0583 & 102.505 & 2356.47  & 0.213  \\ \hline
G3         & 0.0597 & 106.383 & 2377.13  & 0.214  \\ \hline
IOPB-I     & 0.0593 & 105.178 & 2343.66  & 0.212  \\ \hline
IOPB-I+DM3 & 0.0597 & 106.429 & 2387.09  & 0.215   \\ \hline
IOPB-I+DM5 & 0.0603 & 108.224 & 2365.90  & 0.214  \\ \hline  \hline
\end{tabular}
\end{table}
\begin{figure}
\centering
\includegraphics[width=0.5\textwidth]{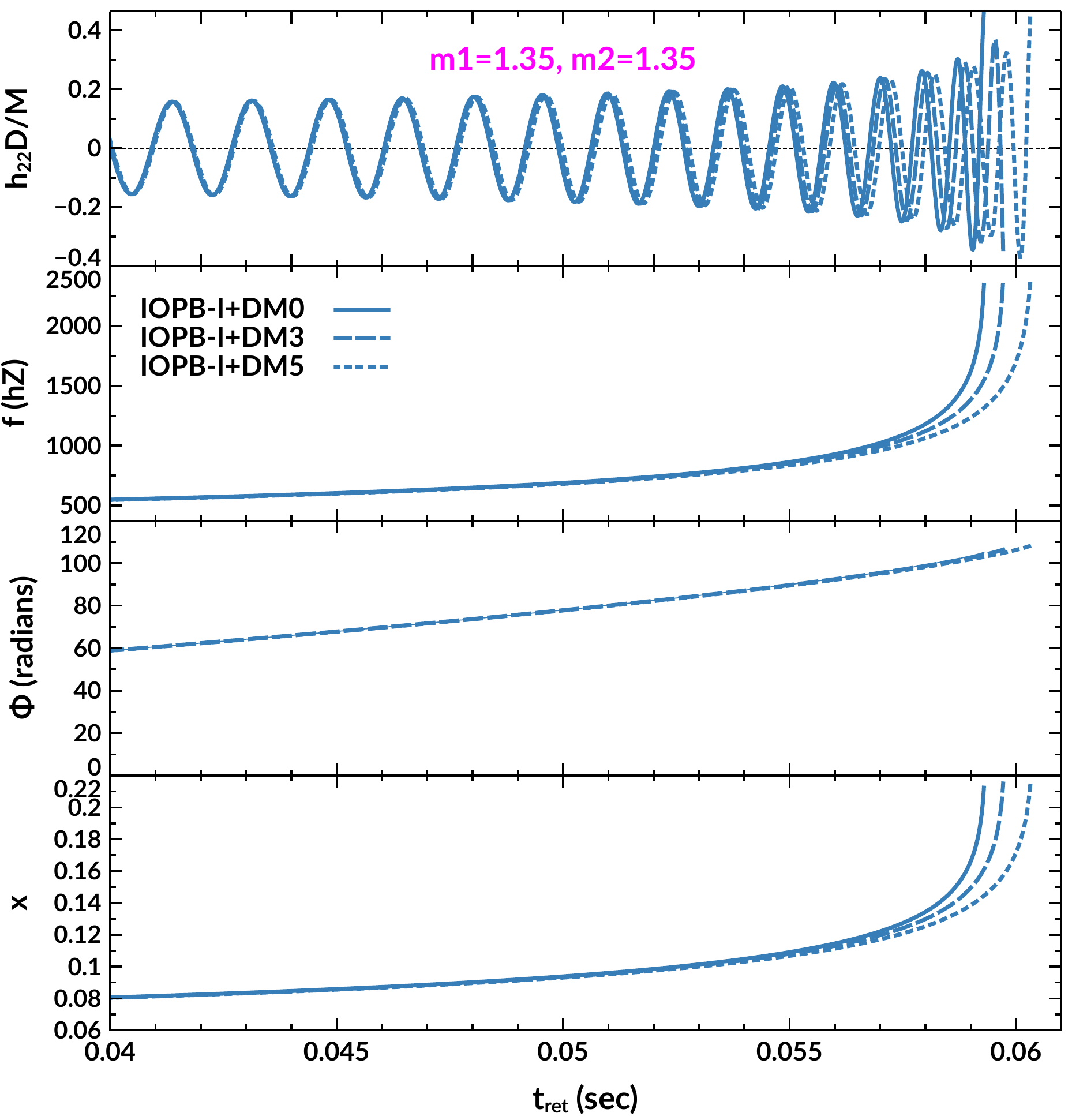}
\caption{(colour online) Same as Fig. \ref{all1}, but for DM admixed NS with IOPB-I parameter set as a representative case.}
\label{all2}
\end{figure}

With the addition of DM to the NS EoS, the calculated macroscopic properties ($M$, $R$ and $\lambda$) decreases. As a result, the tidal interaction between the two NSs becomes weaker as compared to the normal BNS combination. For example, $M=2.149 M_\odot$ without DM and it is 2.051 $M_{\odot}$ with $k_f^{DM}=0.03$ GeV with IOPB-I set. Similarly, the radius of the star is $R_{1.35}=13.168$ km without DM, and it is 12.047 km when DM is included (see Table \ref{table1}). The calculated values for $f$, $\Phi$ and $x$ are tabulated in Table \ref{table2} for IOPB-I parameter set as a representative case with DM momenta $k_f^{DM}=0.03$ and $0.05$ GeV. The time period $t_{ret}$ is found to be 0.0603 sec. with IOPB-I+DM5, which is larger than the other two values obtained with IOPB-I and IOPB-I+DM3. These results infer the more sustainable time in the in-spiral orbits of the BNS system in the presence of DM particles (see Fig. \ref{all2}). The other quantities such as $f$ and $\Phi$ also affected by DM, as shown in Fig. \ref{all2}. 

In general, with the addition of DM, the EOS becomes softer, which affects the macroscopic properties of the BNS, i.e. the mass of the NS decreases with DM. However, a lot of ambiguity prevails in the post-merger remnants of BNS. This has been settled partially after the analysis of GW170817 and GW190425 events. Some of the possible pathways can be found in the Ref. \citep{Sarin_2021}. The likely theoretical probabilities for a post-merger NS remnant basically depend on (i) the remnant mass ($M_R$), and (ii) the EOS of the NS which dictates TOV mass ($M_{TOV}$);
\begin{enumerate}
    \item if $M_R$ $\gtrsim 1.5 \ M_{TOV}$, the system promptly collapses to a black hole.
    \item  for $1.2 \ M_{TOV} \lesssim M \lesssim 1.5 \ M_{TOV}$, a hypermassive neutron star survives the collision, but collapses to form a black hole on dynamical timescales.
    \item in the case, $M_{TOV}<M_R \lesssim 1.2 \ M_{TOV}$, a supramassive NS is survived the collision and will collapses to form a black hole on secular timescales.
    \item if $M_R\leq M_{TOV}$, a stable NS is formed after the merger.
\end{enumerate}
We refer the readers to see Fig. 1 of this Ref. \citep{Sarin_2021} for a clear picture.

The post-merger remnant of DM admixed BNS merger depends on the percentage of DM inside the NS. In Ref. \citep{EllisPLB_2018}, it is reported that for a contribution of DM $\sim$ 5-10\% of the NS mass, an extra peak in the power spectral density of the GW signal is found. Also, for a self-interacting DM of about $\lesssim10\%$, leads to the early formation of a black hole \citep{Pollack_2015}, which can be subsequent seed for supermassive black hole. Therefore, to understand the nature of post-merger remnant of DM admixed BNS, one has to know (i) the exact nature of DM particles and (ii) the percentage of DM present inside the NS.
\section{SUMMARY AND CONCLUSIONS}
\label{con}
In summary, we study the effects of DM of the EMBNS system in the in-spiral phase. The RMF model is used to calculate the EoSs which are input to the TOV equations. In our calculations, the relatively old NL3 parameter set and the two recently developed IOPB-I and G3 forces are used to see the model-dependent nature of the BNS properties. Finally, the calculations are repeated with and without dark matter contain to see the effects at various DM percentage. 

The waveforms of the BNS merger in their in-spiral phase are compared with and without tidal interactions. We find that if the tidal interactions are switched off, the BNS sustain more time in their in-spiral phases. The polarization waveform $h_+$ and $h_\times$ and the strain amplitude of $h_{22}$ mode are calculated for the three-parameter sets. The calculations are done in the retarded time interval at a  distance of 100 Mpc between the source and the observer. Here, we find that the binary system favours a  longer time for softer EoS (G3) as compared to the stiffer NL3 case. The other quantities (frequencies and phase) are almost similar to all the three-parameter sets. However, the in-spiral times in the last orbit are force-dependent.

The DM affects the NS's properties by assuming that the nucleons interact with DM through SM Higgs. The coupling parameters are constrained from both direct detection experiments and LHC searches. The macroscopic properties such as mass, radius, and tidal deformability decrease with the increase of DM percentage inside the NS. The in-spiralling properties are also studied for the DM admixed NS. We find that with the addition of DM, the binary system becomes less deformed, and it sustains more time in their in-spiral phases. The other properties are also affected by the DM as compared to the ordinary NS. Hence, the DM effects on the in-spiral properties of the BNS is significant. Thus, we suggest that the people have to take DM inside the compact objects when they modelling the in-spiral waveforms for the BNS systems.
\section*{Acknowledgement}
HCD would like to thanks Dr. Bharat Kumar and Dr. S. K. Biswal for the discussion on the post-Newtonian formalism and waveform of the gravitational waves.
\section*{DATA AVAILABILITY}
This manuscript has no associated data or the data will not be deposited. Data sharing not applicable to this article as no data sets were generated or analysed during the current study.
\bibliography{pol}

\begin{thebibliography}{}
\makeatletter
\relax
\def\mn@urlcharsother{\let\do\@makeother \do\$\do\&\do\#\do\^\do\_\do\%\do\~}
\def\mn@doi{\begingroup\mn@urlcharsother \@ifnextchar [ {\mn@doi@}
  {\mn@doi@[]}}
\def\mn@doi@[#1]#2{\def\@tempa{#1}\ifx\@tempa\@empty \href
  {http://dx.doi.org/#2} {doi:#2}\else \href {http://dx.doi.org/#2} {#1}\fi
  \endgroup}
\def\mn@eprint#1#2{\mn@eprint@#1:#2::\@nil}
\def\mn@eprint@arXiv#1{\href {http://arxiv.org/abs/#1} {{\tt arXiv:#1}}}
\def\mn@eprint@dblp#1{\href {http://dblp.uni-trier.de/rec/bibtex/#1.xml}
  {dblp:#1}}
\def\mn@eprint@#1:#2:#3:#4\@nil{\def\@tempa {#1}\def\@tempb {#2}\def\@tempc
  {#3}\ifx \@tempc \@empty \let \@tempc \@tempb \let \@tempb \@tempa \fi \ifx
  \@tempb \@empty \def\@tempb {arXiv}\fi \@ifundefined
  {mn@eprint@\@tempb}{\@tempb:\@tempc}{\expandafter \expandafter \csname
  mn@eprint@\@tempb\endcsname \expandafter{\@tempc}}}

\bibitem[\protect\citeauthoryear{Aad et~al.,}{Aad et~al.}{2015}]{Aad_2015}
Aad G.,  et~al., 2015, \mn@doi [Journal of High Energy Physics]
  {10.1007/jhep11(2015)206}, 2015

\bibitem[\protect\citeauthoryear{Abbott, Abbott, Abbott  et~al.}{Abbott
  et~al.}{2017a}]{Abbott_2017}
Abbott B.~P.,  Abbott R.,  Abbott T.~D.,   et~al., 2017a, \mn@doi [Phys. Rev.
  Lett.] {10.1103/PhysRevLett.119.161101}, 119, 161101

\bibitem[\protect\citeauthoryear{Abbott, Abbott, Abbott  et~al.}{Abbott
  et~al.}{2017b}]{Abbott_multi1_2017}
Abbott B.~P.,  Abbott R.,  Abbott T.~D.,   et~al., 2017b, \mn@doi [The
  Astrophysical Journal] {10.3847/2041-8213/aa91c9}, 848, L12

\bibitem[\protect\citeauthoryear{Abbott et~al.,}{Abbott
  et~al.}{2017c}]{Abbott_multi2_2017}
Abbott B.~P.,  et~al., 2017c, \mn@doi [The Astrophysical Journal]
  {10.3847/2041-8213/aa920c}, 848, L13

\bibitem[\protect\citeauthoryear{Abbott, Abbott, Abbott  et~al.}{Abbott
  et~al.}{2018}]{Abbott_2018}
Abbott B.~P.,  Abbott R.,  Abbott T.~D.,   et~al., 2018, \mn@doi [Phys. Rev.
  Lett.] {10.1103/PhysRevLett.121.161101}, 121, 161101

\bibitem[\protect\citeauthoryear{Agathos, Meidam, Del~Pozzo, Li, Tompitak,
  Veitch, Vitale  \& Van Den~Broeck}{Agathos et~al.}{2015}]{Agathos_2015}
Agathos M.,  Meidam J.,  Del~Pozzo W.,  Li T. G.~F.,  Tompitak M.,  Veitch J.,
  Vitale S.,   Van Den~Broeck C.,  2015, \mn@doi [Phys. Rev. D]
  {10.1103/PhysRevD.92.023012}, 92, 023012

\bibitem[\protect\citeauthoryear{Akerib, Alsum, Ara\'ujo  et~al.}{Akerib
  et~al.}{2017}]{LUX_2017}
Akerib D.~S.,  Alsum S.,  Ara\'ujo H.~M.,   et~al., 2017, \mn@doi [Phys. Rev.
  Lett.] {10.1103/PhysRevLett.118.021303}, 118, 021303

\bibitem[\protect\citeauthoryear{Annala, Gorda, Kurkela  \& Vuorinen}{Annala
  et~al.}{2018}]{Annala_2018}
Annala E.,  Gorda T.,  Kurkela A.,   Vuorinen A.,  2018, \mn@doi [Phys. Rev.
  Lett.] {10.1103/PhysRevLett.120.172703}, 120, 172703

\bibitem[\protect\citeauthoryear{Antoniadis, Freire, Wex  et~al.}{Antoniadis
  et~al.}{2013}]{Antoniadis_2013}
Antoniadis J.,  Freire P. C.~C.,  Wex N.,   et~al., 2013, \mn@doi [Science]
  {10.1126/science.1233232}, 340

\bibitem[\protect\citeauthoryear{Aprile et~al.,}{Aprile
  et~al.}{2016}]{Xenon1T_2016}
Aprile E.,  et~al., 2016, \mn@doi [Journal of Cosmology and Astroparticle
  Physics] {10.1088/1475-7516/2016/04/027}, 2016, 027–027

\bibitem[\protect\citeauthoryear{Arun, Blanchet, Iyer  \& Qusailah}{Arun
  et~al.}{2004}]{Arun_2004}
Arun K.~G.,  Blanchet L.,  Iyer B.~R.,   Qusailah M. S.~S.,  2004, \mn@doi
  [Classical and Quantum Gravity] {10.1088/0264-9381/21/15/010}, 21, 3771

\bibitem[\protect\citeauthoryear{Arun, Blanchet, Iyer  \& Qusailah}{Arun
  et~al.}{2005}]{Arun_2005}
Arun K.~G.,  Blanchet L.,  Iyer B.~R.,   Qusailah M. S.~S.,  2005, \mn@doi
  [erratum-ibid] {10.1088/0264-9381/22/14/c01}, 22, 3115

\bibitem[\protect\citeauthoryear{Baiotti, Damour, Giacomazzo, Nagar  \&
  Rezzolla}{Baiotti et~al.}{2011}]{Baiooti_2011}
Baiotti L.,  Damour T.,  Giacomazzo B.,  Nagar A.,   Rezzolla L.,  2011,
  \mn@doi [Phys. Rev. D] {10.1103/PhysRevD.84.024017}, 84, 024017

\bibitem[\protect\citeauthoryear{Bernabei et~al.,}{Bernabei
  et~al.}{2008}]{Bernabei_2008}
Bernabei R.,  et~al., 2008, \mn@doi [EPJ C] {10.1140/epjc/s10052-008-0662-y},
  56, 333–355

\bibitem[\protect\citeauthoryear{Bernal, Heikinheimo, Tenkanen, Tuominen  \&
  Vaskonen}{Bernal et~al.}{2017}]{Bernal_2017}
Bernal N.,  Heikinheimo M.,  Tenkanen T.,  Tuominen K.,   Vaskonen V.,  2017,
  \mn@doi [IJMP A] {10.1142/s0217751x1730023x}, 32, 1730023

\bibitem[\protect\citeauthoryear{Bertone \& Fairbairn}{Bertone \&
  Fairbairn}{2008}]{Bertone_2008}
Bertone G.,  Fairbairn M.,  2008, \mn@doi [Phys. Rev. D]
  {10.1103/PhysRevD.77.043515}, 77, 043515

\bibitem[\protect\citeauthoryear{Bertoni, Nelson  \& Reddy}{Bertoni
  et~al.}{2013}]{Bertoni_2013}
Bertoni B.,  Nelson A.~E.,   Reddy S.,  2013, \mn@doi [Phys. Rev. D]
  {10.1103/PhysRevD.88.123505}, 88, 123505

\bibitem[\protect\citeauthoryear{Bhat \& Paul}{Bhat \& Paul}{2020}]{Bhat_2019}
Bhat S.~A.,  Paul A.,  2020, \mn@doi [The European Physical Journal C]
  {10.1140/epjc/s10052-020-8072-x}, 80, 544

\bibitem[\protect\citeauthoryear{Bini, Damour  \& Faye}{Bini
  et~al.}{2012}]{Bini_2012}
Bini D.,  Damour T.,   Faye G.,  2012, \mn@doi [Phys. Rev. D]
  {10.1103/PhysRevD.85.124034}, 85, 124034

\bibitem[\protect\citeauthoryear{Biswal, Singh, Bhuyan  \& Patra}{Biswal
  et~al.}{2015}]{Biswal_2015}
Biswal S.~K.,  Singh S.~K.,  Bhuyan M.,   Patra S.~K.,  2015, \mn@doi
  [Brazilian Journal of Physics] {10.1007/s13538-015-0317-z}, 45, 347–352

\bibitem[\protect\citeauthoryear{Blaizot}{Blaizot}{1980}]{BLAIZOT_1980}
Blaizot J.,  1980, \mn@doi [Physics Reports]
  {https://doi.org/10.1016/0370-1573(80)90001-0}, 64, 171

\bibitem[\protect\citeauthoryear{Blanchet}{Blanchet}{2006}]{Blanchet_2006}
Blanchet L.,  2006, \mn@doi [Living Reviews in Relativity]
  {10.12942/lrr-2006-4}, 9, 4

\bibitem[\protect\citeauthoryear{Blanchet, Faye, Iyer  \& Sinha}{Blanchet
  et~al.}{2008}]{Blanchet_2008}
Blanchet L.,  Faye G.,  Iyer B.~R.,   Sinha S.,  2008, \mn@doi [Classical and
  Quantum Gravity] {10.1088/0264-9381/25/16/165003}, 25, 165003

\bibitem[\protect\citeauthoryear{Bodmer}{Bodmer}{1991}]{BODMER_1991}
Bodmer A.,  1991, \mn@doi [Nuclear Physics A]
  {https://doi.org/10.1016/0375-9474(91)90439-D}, 526, 703

\bibitem[\protect\citeauthoryear{Boyle, Brown, Kidder, Mrou\'e, Pfeiffer,
  Scheel, Cook  \& Teukolsky}{Boyle et~al.}{2007}]{Boyle_2007}
Boyle M.,  Brown D.~A.,  Kidder L.~E.,  Mrou\'e A.~H.,  Pfeiffer H.~P.,  Scheel
  M.~A.,  Cook G.~B.,   Teukolsky S.~A.,  2007, \mn@doi [Phys. Rev. D]
  {10.1103/PhysRevD.76.124038}, 76, 124038

\bibitem[\protect\citeauthoryear{Bunta \& Gmuca}{Bunta \&
  Gmuca}{2003}]{Bunta_2003}
Bunta J.~K.,  Gmuca S.,  2003, \mn@doi [Phys. Rev. C]
  {10.1103/PhysRevC.68.054318}, 68, 054318

\bibitem[\protect\citeauthoryear{Buonanno \& Sathyaprakash}{Buonanno \&
  Sathyaprakash}{2014}]{Buonanno_2014}
Buonanno A.,  Sathyaprakash B.~S.,  2014, {Sources of Gravitational Waves:
  Theory and Observations}.
 (\mn@eprint {arXiv} {1410.7832})

\bibitem[\protect\citeauthoryear{{CALTECH}}{{CALTECH}}{2002}]{LIGO}
{CALTECH} 2002, {LIGO Scientific Collaboration}, \url
  {https://www.ligo.caltech.edu}

\bibitem[\protect\citeauthoryear{{CDMS}}{{CDMS}}{2010}]{CDMS_2010}
{CDMS} 2010, \mn@doi [Science] {10.1126/science.1186112}, 327, 1619–1621

\bibitem[\protect\citeauthoryear{Chornock et~al.,}{Chornock
  et~al.}{2017}]{Chornock_2017}
Chornock R.,  et~al., 2017, \mn@doi [The Astrophysical Journal]
  {10.3847/2041-8213/aa905c}, 848, L19

\bibitem[\protect\citeauthoryear{Ciarcelluti \& Sandin}{Ciarcelluti \&
  Sandin}{2011}]{Ciarcelluti_2011}
Ciarcelluti P.,  Sandin F.,  2011, \mn@doi [Phys. Lett. B]
  {10.1016/j.physletb.2010.11.021}, 695, 19–21

\bibitem[\protect\citeauthoryear{Cowperthwaite et~al.,}{Cowperthwaite
  et~al.}{2017}]{Cowperthwaite_2017}
Cowperthwaite P.~S.,  et~al., 2017, \mn@doi [The Astrophysical Journal]
  {10.3847/2041-8213/aa8fc7}, 848, L17

\bibitem[\protect\citeauthoryear{Creighton \& Anderson}{Creighton \&
  Anderson}{2011}]{Creighton_2011}
Creighton J.,  Anderson W.,  2011, Gravitational-Wave Physics and Astronomy: An
  Introduction to Theory, Experiment and Data Analysis.
Wiley Series in Cosmology, Wiley, \url
  {https://books.google.co.in/books?id=jLRLVaR97AUC}

\bibitem[\protect\citeauthoryear{Cromartie, Fonseca, Ransom  et~al.}{Cromartie
  et~al.}{2019}]{Cromartie_2019}
Cromartie H.~T.,  Fonseca E.,  Ransom S.~M.,   et~al., 2019, \mn@doi [Nature
  Astronomy] {10.1038/s41550-019-0880-2}, 4, 72–76

\bibitem[\protect\citeauthoryear{Czarnecki, K\"orner  \& Piclum}{Czarnecki
  et~al.}{2010}]{Czarnecki_2010}
Czarnecki A.,  K\"orner J.~G.,   Piclum J.~H.,  2010, \mn@doi [Phys. Rev. D]
  {10.1103/PhysRevD.81.111503}, 81, 111503

\bibitem[\protect\citeauthoryear{Das, Malik  \& Nayak}{Das
  et~al.}{2019}]{Das_2019}
Das A.,  Malik T.,   Nayak A.~C.,  2019, \mn@doi [Phys. Rev. D]
  {10.1103/PhysRevD.99.043016}, 99, 043016

\bibitem[\protect\citeauthoryear{Das, Kumar, Kumar  et~al.}{Das
  et~al.}{2020}]{Das_2020}
Das H.~C.,  Kumar A.,  Kumar B.,   et~al., 2020, \mn@doi [MNRAS]
  {10.1093/mnras/staa1435}, 495, 4893

\bibitem[\protect\citeauthoryear{Das, Kumar, Kumar, Biswal  \& Patra}{Das
  et~al.}{2021}]{Das_2021}
Das H.~C.,  Kumar A.,  Kumar B.,  Biswal S.~K.,   Patra S.~K.,  2021, \mn@doi
  [Journal of Cosmology and Astroparticle Physics]
  {10.1088/1475-7516/2021/01/007}, 2021, 007

\bibitem[\protect\citeauthoryear{De~Lavallaz \& Fairbairn}{De~Lavallaz \&
  Fairbairn}{2010}]{De_Lavallaz_2010}
De~Lavallaz A.,  Fairbairn M.,  2010, \mn@doi [Phys. Rev. D]
  {10.1103/PhysRevD.81.123521}, 81, 123521

\bibitem[\protect\citeauthoryear{Djouadi, Lebedev, Mambrini  \&
  Quevillon}{Djouadi et~al.}{2012}]{Djouadi_2012}
Djouadi A.,  Lebedev O.,  Mambrini Y.,   Quevillon J.,  2012, \mn@doi [Physics
  Letters B] {https://doi.org/10.1016/j.physletb.2012.01.062}, 709, 65

\bibitem[\protect\citeauthoryear{Duffy \& van Bibber}{Duffy \& van
  Bibber}{2009}]{Duffy_2009}
Duffy L.~D.,  van Bibber K.,  2009, \mn@doi [New Journal of Physics]
  {10.1088/1367-2630/11/10/105008}, 11, 105008

\bibitem[\protect\citeauthoryear{Dutra, Louren\ifmmode~\mbox{\c{c}}\else
  \c{c}\fi{}o, S\'a~Martins, Delfino, Stone  \& Stevenson}{Dutra
  et~al.}{2012}]{Dutra_2012}
Dutra M.,  Louren\ifmmode~\mbox{\c{c}}\else \c{c}\fi{}o O.,  S\'a~Martins
  J.~S.,  Delfino A.,  Stone J.~R.,   Stevenson P.~D.,  2012, \mn@doi [Phys.
  Rev. C] {10.1103/PhysRevC.85.035201}, 85, 035201

\bibitem[\protect\citeauthoryear{Ellis, Hektor, Hütsi, Kannike, Marzola,
  Raidal  \& Vaskonen}{Ellis et~al.}{2018}]{EllisPLB_2018}
Ellis J.,  Hektor A.,  Hütsi G.,  Kannike K.,  Marzola L.,  Raidal M.,
  Vaskonen V.,  2018, \mn@doi [Physics Letters B]
  {https://doi.org/10.1016/j.physletb.2018.04.048}, 781, 607

\bibitem[\protect\citeauthoryear{Favata}{Favata}{2014}]{Favata_2014}
Favata M.,  2014, \mn@doi [Phys. Rev. Lett.] {10.1103/PhysRevLett.112.101101},
  112, 101101

\bibitem[\protect\citeauthoryear{Flanagan \& Hinderer}{Flanagan \&
  Hinderer}{2008}]{Flangnan_2008}
Flanagan E.~E.,  Hinderer T.,  2008, \mn@doi [Phys. Rev. D]
  {10.1103/PhysRevD.77.021502}, 77, 021502

\bibitem[\protect\citeauthoryear{Gmuca}{Gmuca}{1992}]{GMUCA_1992}
Gmuca S.,  1992, \mn@doi [Nuclear Physics A]
  {https://doi.org/10.1016/0375-9474(92)90032-F}, 547, 447

\bibitem[\protect\citeauthoryear{Goldstein et~al.,}{Goldstein
  et~al.}{2017}]{Goldstein_grb_2017}
Goldstein A.,  et~al., 2017, \mn@doi [The Astrophysical Journal]
  {10.3847/2041-8213/aa8f41}, 848, L14

\bibitem[\protect\citeauthoryear{Hall, Jedamzik, March-Russell  \& West}{Hall
  et~al.}{2010}]{Hall_2010}
Hall L.~J.,  Jedamzik K.,  March-Russell J.,   West S.~M.,  2010, \mn@doi
  [JHEP] {10.1007/jhep03(2010)080}, 2010

\bibitem[\protect\citeauthoryear{Hallinan et~al.,}{Hallinan
  et~al.}{2017}]{Hallinan_2017}
Hallinan G.,  et~al., 2017, \mn@doi [Science] {10.1126/science.aap9855}, 358,
  1579–1583

\bibitem[\protect\citeauthoryear{Hinderer}{Hinderer}{2008}]{Hinderer_2008}
Hinderer T.,  2008, \mn@doi [The Astrophysical Journal] {10.1086/533487}, 677,
  1216

\bibitem[\protect\citeauthoryear{Hinderer, Lackey, Lang  \& Read}{Hinderer
  et~al.}{2010}]{Hinderer_2010}
Hinderer T.,  Lackey B.~D.,  Lang R.~N.,   Read J.~S.,  2010, \mn@doi [Phys.
  Rev. D] {10.1103/PhysRevD.81.123016}, 81, 123016

\bibitem[\protect\citeauthoryear{Hotokezaka, Kyutoku  \& Shibata}{Hotokezaka
  et~al.}{2013}]{Hotokezaka_2013}
Hotokezaka K.,  Kyutoku K.,   Shibata M.,  2013, \mn@doi [Phys. Rev. D]
  {10.1103/PhysRevD.87.044001}, 87, 044001

\bibitem[\protect\citeauthoryear{Hotokezaka, Kyutoku, Sekiguchi  \&
  Shibata}{Hotokezaka et~al.}{2016}]{Hotokezaka_2016}
Hotokezaka K.,  Kyutoku K.,  Sekiguchi Y.-i.,   Shibata M.,  2016, \mn@doi
  [Phys. Rev. D] {10.1103/PhysRevD.93.064082}, 93, 064082

\bibitem[\protect\citeauthoryear{{INFN}}{{INFN}}{1993}]{VIRGO}
{INFN} 1993, {VIRGO Project Central Web site}, \url {http://www.virgo.infn.it.}

\bibitem[\protect\citeauthoryear{Isoyama, Sturani  \& Nakano}{Isoyama
  et~al.}{2020}]{Isoyama_2020}
Isoyama S.,  Sturani R.,   Nakano H.,  2020, {Post-Newtonian templates for
  gravitational waves from compact binary inspirals} (\mn@eprint {arXiv}
  {2012.01350})

\bibitem[\protect\citeauthoryear{{KAGRA}}{{KAGRA}}{2010}]{KAGRA}
{KAGRA} 2010, {Kagra Observatory}, \url
  {https://gwcenter.icrr.u-tokyo.ac.jp/en/}

\bibitem[\protect\citeauthoryear{Kidder}{Kidder}{2008}]{Kidder_2008}
Kidder L.~E.,  2008, \mn@doi [Phys. Rev. D] {10.1103/PhysRevD.77.044016}, 77,
  044016

\bibitem[\protect\citeauthoryear{Kidder, Blanchet  \& Iyer}{Kidder
  et~al.}{2007}]{Kidder_2007}
Kidder L.~E.,  Blanchet L.,   Iyer B.~R.,  2007, \mn@doi [Classical and Quantum
  Gravity] {10.1088/0264-9381/24/20/n01}, 24, 5307

\bibitem[\protect\citeauthoryear{Kouvaris}{Kouvaris}{2008}]{Kouvaris_2008}
Kouvaris C.,  2008, \mn@doi [Phys. Rev. D] {10.1103/PhysRevD.77.023006}, 77,
  023006

\bibitem[\protect\citeauthoryear{Kouvaris \& Tinyakov}{Kouvaris \&
  Tinyakov}{2010}]{Kouvaris_2010}
Kouvaris C.,  Tinyakov P.,  2010, \mn@doi [Phys. Rev. D]
  {10.1103/PhysRevD.82.063531}, 82, 063531

\bibitem[\protect\citeauthoryear{Kouvaris \& Tinyakov}{Kouvaris \&
  Tinyakov}{2011}]{Kouvaris_2011}
Kouvaris C.,  Tinyakov P.,  2011, \mn@doi [Phys. Rev. D]
  {10.1103/PhysRevD.83.083512}, 83, 083512

\bibitem[\protect\citeauthoryear{Kumar, Singh, Agrawal  \& Patra}{Kumar
  et~al.}{2017}]{Kumar_2017}
Kumar B.,  Singh S.,  Agrawal B.,   Patra S.,  2017, \mn@doi [Nuclear Physics
  A] {https://doi.org/10.1016/j.nuclphysa.2017.07.001}, 966, 197

\bibitem[\protect\citeauthoryear{Kumar, Patra  \& Agrawal}{Kumar
  et~al.}{2018}]{Kumar_2018}
Kumar B.,  Patra S.~K.,   Agrawal B.~K.,  2018, \mn@doi [Phys. Rev. C]
  {10.1103/PhysRevC.97.045806}, 97, 045806

\bibitem[\protect\citeauthoryear{Lackey}{Lackey}{2012}]{Lackey_2012}
Lackey B.~D.,  2012, PhD thesis, \url {https://dc.uwm.edu/etd/72}

\bibitem[\protect\citeauthoryear{Lalazissis, K\"onig  \& Ring}{Lalazissis
  et~al.}{1997}]{Lalazissis_1997}
Lalazissis G.~A.,  K\"onig J.,   Ring P.,  1997, \mn@doi [Phys. Rev. C]
  {10.1103/PhysRevC.55.540}, 55, 540

\bibitem[\protect\citeauthoryear{MARTIN}{MARTIN}{}]{Martin_1998}
MARTIN S.~P., , A SUPERSYMMETRY PRIMER, \mn@doi{10.1142/9789812839657_0001.
}, \url {https://www.worldscientific.com/doi/abs/10.1142/9789812839657_0001}

\bibitem[\protect\citeauthoryear{Maggiore}{Maggiore}{2008}]{Maggiore_2008}
Maggiore M.,  2008, Gravitational Waves: Volume 1: Theory and Experiments.
Gravitational Waves, OUP Oxford, \url
  {https://books.google.co.in/books?id=AqVpQgAACAAJ}

\bibitem[\protect\citeauthoryear{Margalit \& Metzger}{Margalit \&
  Metzger}{2017}]{Margalit_2017}
Margalit B.,  Metzger B.~D.,  2017, \mn@doi [The Astrophysical Journal]
  {10.3847/2041-8213/aa991c}, 850, L19

\bibitem[\protect\citeauthoryear{Margutti et~al.,}{Margutti
  et~al.}{2017}]{Margutti_2017}
Margutti R.,  et~al., 2017, \mn@doi [The Astrophysical Journal]
  {10.3847/2041-8213/aa9057}, 848, L20

\bibitem[\protect\citeauthoryear{Miller, Lamb, Dittmann  et~al.}{Miller
  et~al.}{2019}]{Miller_2019}
Miller M.~C.,  Lamb F.~K.,  Dittmann A.~J.,   et~al., 2019, \mn@doi [APJ]
  {10.3847/2041-8213/ab50c5}, 887, L24

\bibitem[\protect\citeauthoryear{Most, Weih, Rezzolla  \&
  Schaffner-Bielich}{Most et~al.}{2018}]{Most_2018}
Most E.~R.,  Weih L.~R.,  Rezzolla L.,   Schaffner-Bielich J.,  2018, \mn@doi
  [Phys. Rev. Lett.] {10.1103/PhysRevLett.120.261103}, 120, 261103

\bibitem[\protect\citeauthoryear{Nandi, Char  \& Pal}{Nandi
  et~al.}{2019}]{Nandi_2019}
Nandi R.,  Char P.,   Pal S.,  2019, \mn@doi [Phys. Rev. C]
  {10.1103/PhysRevC.99.052802}, 99, 052802

\bibitem[\protect\citeauthoryear{Nicholl et~al.,}{Nicholl
  et~al.}{2017}]{Nicholl_2017}
Nicholl M.,  et~al., 2017, \mn@doi [The Astrophysical Journal]
  {10.3847/2041-8213/aa9029}, 848, L18

\bibitem[\protect\citeauthoryear{Panotopoulos \& Lopes}{Panotopoulos \&
  Lopes}{2017}]{Panotopoulos_2017}
Panotopoulos G.,  Lopes I.,  2017, \mn@doi [Phys. Rev. D]
  {10.1103/PhysRevD.96.083004}, 96, 083004

\bibitem[\protect\citeauthoryear{Pollack, Spergel  \& Steinhardt}{Pollack
  et~al.}{2015}]{Pollack_2015}
Pollack J.,  Spergel D.~N.,   Steinhardt P.~J.,  2015, \mn@doi [The
  Astrophysical Journal] {10.1088/0004-637x/804/2/131}, 804, 131

\bibitem[\protect\citeauthoryear{Quddus, Panotopoulos, Kumar, Ahmad  \&
  Patra}{Quddus et~al.}{2020}]{Quddus_2020}
Quddus A.,  Panotopoulos G.,  Kumar B.,  Ahmad S.,   Patra S.~K.,  2020,
  \mn@doi [Journal of Physics G: Nuclear and Particle Physics]
  {10.1088/1361-6471/ab9d36}, 47, 095202

\bibitem[\protect\citeauthoryear{Radice, Perego, Zappa  \& Bernuzzi}{Radice
  et~al.}{2018}]{Radice_2018}
Radice D.,  Perego A.,  Zappa F.,   Bernuzzi S.,  2018, \mn@doi [The
  Astrophysical Journal] {10.3847/2041-8213/aaa402}, 852, L29

\bibitem[\protect\citeauthoryear{Read et~al.,}{Read et~al.}{2013}]{Read_2013}
Read J.~S.,  et~al., 2013, \mn@doi [Phys. Rev. D] {10.1103/PhysRevD.88.044042},
  88, 044042

\bibitem[\protect\citeauthoryear{Riley, Watts, Bogdanov  et~al.}{Riley
  et~al.}{2019}]{Riley_2019}
Riley T.~E.,  Watts A.~L.,  Bogdanov S.,   et~al., 2019, \mn@doi [APJ]
  {10.3847/2041-8213/ab481c}, 887, L21

\bibitem[\protect\citeauthoryear{Ruppin, Billard, Figueroa-Feliciano  \&
  Strigari}{Ruppin et~al.}{2014}]{Ruppin_2014}
Ruppin F.,  Billard J.,  Figueroa-Feliciano E.,   Strigari L.,  2014, \mn@doi
  [Phys. Rev. D] {10.1103/PhysRevD.90.083510}, 90, 083510

\bibitem[\protect\citeauthoryear{Sarin \& Lasky}{Sarin \&
  Lasky}{2021}]{Sarin_2021}
Sarin N.,  Lasky P.~D.,  2021, \mn@doi [General Relativity and Gravitation]
  {10.1007/s10714-021-02831-1}, 53, 59

\bibitem[\protect\citeauthoryear{Sharma, Centelles, Vi\~nas, Baldo  \&
  Burgio}{Sharma et~al.}{2015}]{BKS_2015}
Sharma B.~K.,  Centelles M.,  Vi\~nas X.,  Baldo M.,   Burgio G.~F.,  2015,
  \mn@doi [A\&A] {10.1051/0004-6361/201526642}, 584, A103

\bibitem[\protect\citeauthoryear{Sugahara \& Toki}{Sugahara \&
  Toki}{1994}]{Toki_1994}
Sugahara Y.,  Toki H.,  1994, \mn@doi [Nuclear Physics A]
  {https://doi.org/10.1016/0375-9474(94)90923-7}, 579, 557

\bibitem[\protect\citeauthoryear{Tan, Xiao, Cui  et~al.}{Tan
  et~al.}{2016}]{PandaX_2016}
Tan A.,  Xiao M.,  Cui X.,   et~al., 2016, \mn@doi [Phys. Rev. Lett.]
  {10.1103/PhysRevLett.117.121303}, 117, 121303

\bibitem[\protect\citeauthoryear{Toussaint \& Freeman}{Toussaint \&
  Freeman}{2009}]{MILC_2009}
Toussaint D.,  Freeman W.,  2009, \mn@doi [Phys. Rev. Lett.]
  {10.1103/PhysRevLett.103.122002}, 103, 122002

\bibitem[\protect\citeauthoryear{Troja et~al.,}{Troja
  et~al.}{2017}]{Troja_2017}
Troja E.,  et~al., 2017, \mn@doi [Nature] {10.1038/nature24290}, 551, 71–74

\bibitem[\protect\citeauthoryear{Vines, Flanagan  \& Hinderer}{Vines
  et~al.}{2011}]{Vines_2011}
Vines J.,  Flanagan E.~E.,   Hinderer T.,  2011, \mn@doi [Phys. Rev. D]
  {10.1103/PhysRevD.83.084051}, 83, 084051

\bibitem[\protect\citeauthoryear{Wade, Creighton, Ochsner, Lackey, Farr,
  Littenberg  \& Raymond}{Wade et~al.}{2014}]{Wade_2014}
Wade L.,  Creighton J. D.~E.,  Ochsner E.,  Lackey B.~D.,  Farr B.~F.,
  Littenberg T.~B.,   Raymond V.,  2014, \mn@doi [Phys. Rev. D]
  {10.1103/PhysRevD.89.103012}, 89, 103012

\bibitem[\protect\citeauthoryear{Zhang, Li  \& Xu}{Zhang
  et~al.}{2018}]{Zhang_2018}
Zhang N.-B.,  Li B.-A.,   Xu J.,  2018, \mn@doi [The Astrophysical Journal]
  {10.3847/1538-4357/aac027}, 859, 90

\makeatother
\end{thebibliography}
\bibliographystyle{mnras}
\end{document}